\documentclass[preprint]{revtex4}

\usepackage[dvips]{color}
\usepackage{amsmath}
\usepackage{graphicx}
\usepackage{amssymb}
\usepackage{amsthm, amscd}
\usepackage{amsfonts}
\usepackage{cancel}
\usepackage{times}
\usepackage{pstricks}
\usepackage{wasysym}
\usepackage{ulem}
\usepackage[dvips]{epsfig}
\usepackage{subfigure}

\newcommand{\dslash}{\not{\hbox{\kern-2pt $\partial$}}}
\newcommand{\bq}{\begin{equation}} 
\newcommand{\eq}{\end{equation}}
\newcommand{\bqa}{\begin{eqnarray}} 
\newcommand{\eqa}{\end{eqnarray}}
\newcommand{\nn}{\nonumber \\}
\newcommand{\bw}{\begin{widetext}}
\newcommand{\ew}{\end{widetext}}

\begin{document}


\title{A Non-Fermi Liquid from a Charged Black Hole; A Critical Fermi Ball}

\author{Sung-Sik Lee}
\affiliation{Department of Physics $\&$ Astronomy, 
McMaster University,
Hamilton, Ontario L8S 4M1, 
Canada}

\date{\today}

\begin{abstract}
Using the AdS/CFT correspondence, we calculate a fermionic spectral function in a 2+1 dimensional non-relativistic quantum field theory which is dual to a gravitational theory in the $AdS_4$ background with a charged black hole.
The spectral function shows no quasiparticle peak but the Fermi surface is still well defined.
Interestingly, all momentum points inside the Fermi surface are critical and 
the gapless modes are defined in a {\it critical Fermi ball} in the momentum space.

\end{abstract}

\maketitle


Landau Fermi liquid theory is a low energy effective theory for degenerate many-body Fermi systems.
In Fermi liquid states, 
the curvature of a Fermi surface suppresses non-forward scatterings, 
which makes it possible for well defined quasiparticles to exist in the low energy limit\cite{SHANKAR,POLCHINSKI}.
Then non-interacting Fermi gases essentially capture the qualitative nature of interacting Fermi systems.
Although the Fermi liquid theory provides correct descriptions for most metals,
strongly correlated systems including the high temperature superconductors\cite{PALEE} and heavy fermion compounds\cite{LOHNEYSEN,GEGENWART} are not described by the Fermi liquid theory.
These so called non-Fermi liquid states are different states of conducting matters 
and an understanding of those states is very important
not only for applications but also from a fundamental physics point of view.
However, theoretical understandings of non-Fermi liquid states are still limited.
In particular, there is no general theoretical tool 
which enables one to make quantitative predictions.
The fundamental difficulty lies in the fact that for strongly correlated non-Fermi liquid states the well developed standard perturbation theories usually break down.

On the other hand, there have been significant developments in understanding 
a class of strongly coupled quantum field theories.
In the AdS/CFT correspondence, 
a gravitational theory in the anti-de Sitter (AdS) space
is dual to a strongly coupled conformal field theory (CFT) 
defined on the boundary of the AdS space\cite{MALDACENA,GUBSER,WITTEN}.
In a large N limit, the gravitational theory is reduced to a classical gravity from which
one can understand non-trivial strong coupling physics of the boundary CFT.
Recently the AdS/CFT correspondence has been applied to various phenomena 
which arise in the context of condensed matter systems\cite{HARTNOLL,HARTNOLL_NERNST,HERZOG,KARCH,GUBSER_SC,HARTNOLL_SC,SON,KARCH_ZS}.
Therefore it is of interest to find a dual gravitational description for a non-Fermi liquid state.
The goal of this paper is to study dynamical properties of a strongly interacting non-relativistic quantum field theory by calculating a fermionic spectral function from the AdS/CFT correspondence.
For this, we consider a gravitational background of a charged black hole where 
the charge induces a non-zero density of fermions in the boundary quantum field theory.

We consider the action,
\bqa
&& S  =  \frac{1}{\kappa_4^2} \int d^4 x \sqrt{-g} \left[ \frac{1}{4}R  - \frac{1}{4}F_{\mu \nu} F^{\mu \nu} + \frac{3}{2} \right] \nn
&& + \int d^4 x \sqrt{-g} \left[ \bar \psi \gamma^\mu \left( \partial_\mu + \frac{1}{2} \omega^{bc}_\mu \Sigma_{bc} - i A_\mu \right) \psi - m \bar \psi \psi \right] \nn
&& + \int d^3 x \sqrt{- g_\epsilon} \bar \psi \psi.
\label{action}
\eqa
Here $x^\mu = (t,x,y,z)$ is the space-time coordinate with signature $(-1,1,1,1)$.
$R$ is the scalar curvature and $F_{\mu \nu} = \partial_\mu A_\nu - \partial_\nu A_\mu$ is the field strength tensor of a U(1) gauge field $A_\mu$.
$\psi$ is a four-component Dirac spinor, $\omega^{bc}_\mu$ is the spin connection and 
$\Sigma_{bc} = \frac{1}{4} [ \Gamma_b, \Gamma_c ]$ is the generator of the local Lorentz transformation with $\Gamma_a$, the gamma matrices.
$\gamma^\mu = e^\mu_a \Gamma^a$ where $e^\mu_a$ is the tetrad.
The action describe the U(1) gauge field and the Dirac spinor coupled with gravity in the background with a negative cosmological constant which is set to be $-1$ in our unit.
The last term in the action is a boundary term defined at $z=\epsilon$ where $g_\epsilon$ is the determinant of the induced metric on the 2+1D space.
Although the boundary term does not affect the equation of motion in the bulk,
it is important for obtaining a non-trivial dependence of the saddle point action 
on the boundary value of the spinor field\cite{HENNINGSON}.

The above action has an $AdS_4$ black hole solution given by
\bqa
&& ds^2 = \frac{1}{z^2} \left[
\alpha^2 ( -f(z)^2 dt^2 + dx^2 + dy^2 ) + \frac{dz^2}{f(z)^2} \right], \nn
&& A_0 = q \alpha (z-1),
\eqa
where $f(z) = \sqrt{ 1 + q^2 z^4 - (1+q^2) z^3 }$.
This is a special case of a more general dyonic black hole solution considered in Ref. \cite{HARTNOLL}.
In this coordinate system, the horizon of the black hole is at $z=1$.
The metric describes the $AdS_4$ space near the boundary at $z=0$.
The Hawking temperature of the black hole is $T_H = \frac{\alpha}{4\pi} ( 3 - q^2)$\cite{HARTNOLL}.
For nonzero $q$ and $\alpha$, the black hole carries a nonzero charge.
Non-vanishing components of the spin connection are 
$\omega^{\hat t \hat z}_{\hat t} = z^2 \left( \frac{f}{z} \right)^{'}$ and
$\omega^{\hat x \hat z}_{\hat x} = \omega^{\hat y \hat z}_{\hat y} =-f $, where
$(\hat t,\hat x,\hat y,\hat z)$ represents the local Lorentz coordinate.

According to the AdS/CFT correspondence, we can view this classical gravitational theory as 
a strongly coupled 2+1D quantum field theory in a large $N$ limit.
The Hawking temperature corresponds to the temperature of the boundary field theory.
This theory can be motivated from the M-theory 
defined in $AdS_4 \times S^7$ which describes
the low energy physics of the 2+1D supersymmetric Yang-Mills theory 
with $16$ supercharges.
If the infinite tower of the Kaluza-Klein modes are truncated self consistently in the M-theory, 
the resulting theory would contain the above theory with
particular values of the fermionic mass.
Here, we will not restrict ourselves to the M-theory and
we will regard the fermionic mass as a free parameter 
which characterizes the corresponding 2+1D quantum field theory.
In particular, we will focus on the case with $m=0$ where
the  chiral symmetry simplifies the calculation significantly.
However, the qualitative features may be similar for other values of the mass.

Recently, it has been proposed that a 2+1 dimensional $U(N) \times U(N)$ Chern-Simons-matter theory at level $k$ 
is dual to the type IIA string theory on $AdS_4 \times CP^3$ in a 't Hooft limit with a fixed $\lambda = N/k$\cite{ABJM}.
The present gravitational theory may be related to the Chern-Simons-matter theory in the strong coupling limit ($\lambda >> 1$) at a finite chemical potential.
In this paper, instead of attempting to establish a precise connection with a microscopic theory, we take the gravitational theory in Eq. (\ref{action}) as our starting point and examine the dynamics of the fermion, in a hope that the gravity description may capture some universal features of strongly interacting fermions at finite density.

The Dirac spinor is a source field which is linearly coupled with a fermionic field in the boundary theory.
The gauge field is coupled with a conserved U(1) current which includes the current of the fermion. 
The  electrostatic potential induces a nonzero density of the boundary fermions.
The chemical potential of the boundary theory is given by $\mu = A_0(z=0) = -q \alpha$.
This is crucial in obtaining a system of fermions with a finite density.
It is noted that the fermionic field that couples with the Dirac spinor can be a composite field in the ultra-violet theory.
In the following, we will calculate the `single particle' spectral function of the fermion which is possibly a composite particle.

For $m=0$, the chiral symmetry enables us to focus on one chiral mode.
Due to the 2+1 dimensional translational symmetry, 
we can assume a plane wave solution 
for the left chiral modes $\psi_{-}$ and $\bar \psi_-$ which satisfy
$\Gamma^5 \psi_- = - \psi_-$ and $\bar \psi_- \Gamma^5 = - \bar \psi_-$,
\bqa
\psi(t,x,y,z) & = & e^{ - i (\omega t -  {\bf k} \cdot {\bf r} )} \psi_{-}(z), \nn
\bar \psi(t,x,y,z) & = & e^{  i (\omega t -  {\bf k} \cdot {\bf r} )} \bar \psi_{-}(z), 
\eqa
where ${\bf r}=(x,y)$ and ${\bf k}=(k_x,k_y)$.
In the chiral representation with
$\Gamma^{\hat 0} = \left( \begin{array}{cc} 0 & -I \\ I & 0 \end{array} \right)$, 
$\Gamma^{\hat i} = \left( \begin{array}{cc} 0 & \sigma^i \\ \sigma^i & 0 \end{array}\right)$,
where $\sigma^i$ are the Pauli matrices with $i=x,y,z$ 
and $I$ is the $2 \times 2$ identity matrix,
the equation of motion for the two-component chiral spinors becomes
\bqa
\left[
\frac{ i \omega z}{\alpha f} + \frac{i q z (z-1)}{f} + \frac{i {\bf k} \cdot {\bf \sigma} }{\alpha} z  
+ \left(
\frac{z f^{'}}{2} - \frac{3 f}{2} + z f \partial_z 
\right) \sigma^z 
\right] \psi_{-}(z) = 0, \nn
\bar \psi_-(z) 
\left[
 \frac{ i \omega z}{\alpha f} + \frac{i q z (z-1)}{f} - \frac{i {\bf k} \cdot {\bf \sigma} }{\alpha} z  
+ \left(
\frac{z f^{'}}{2} - \frac{3 f}{2} + \overleftarrow{\partial_z} z f  
\right) \sigma^z 
\right] = 0.
\label{eom}
\eqa
At $T=0$ ($q^2=3$), the solution near the horizon ($z \rightarrow 1$) reads
$\psi_-(z) \sim e^{ - i \omega  \frac{1}{6 \alpha (1-z)} \sigma^z } C$, 
$\bar \psi_-(z) \sim \bar C e^{ - i \omega  \frac{1}{6 \alpha (1-z)} \sigma^z }$
in the leading order of $(1-z)$
for some fixed spinors $C$, $\bar C$. 
To maintain the causality in the boundary theory,
we impose the ingoing boundary condition, 
\bqa
\psi_{-}(z) 
 \sim  
e^{ i \omega  \frac{1}{6 \alpha (1-z)} } 
\left( \begin{array}{c} 0 \\ 1 \end{array} \right), ~~
\bar \psi_{-}^T(z) 
 \sim 
e^{ -i \omega  \frac{1}{6 \alpha (1-z)} } 
\left( \begin{array}{c} 1 \\ 0 \end{array} \right)
\label{inBC}
\eqa
as $z \rightarrow 1$.
At a finite temperature with $q^2 < 3$, the boundary condition is modified to be 
$\psi_- \sim e^{ -i \omega  \frac{\ln (1-z) }{\alpha (3-q^2)} } 
\left( \begin{array}{c} 0 \\ 1 \end{array} \right)$ and
$\psi_-^T \sim e^{ i \omega  \frac{\ln (1-z) }{\alpha (3-q^2)} } 
\left( \begin{array}{c} 1 \\ 0 \end{array} \right)$.
Therefore there is one-parameter family of solutions 
that satisfy the ingoing boundary condition near the horizon.
Near the boundary of the AdS space ($z \rightarrow 0$), the spinors behave as 
$\psi_-(z) \sim z^{3/2} \chi$ and 
$\bar \psi_-(z) \sim z^{3/2} \bar \chi$, 
where $\chi$ and $\bar \chi$ should be chosen so that the solution
satisfies the ingoing boundary condition near the horizon.
We represent the solution near the boundary as
\bqa
\psi_-(z) & \sim & \left( \frac{z}{\alpha} \right)^{3/2} \left( \begin{array}{c} P(\omega,{\bf k}) \\ 1 \end{array} \right) \eta, \nn
\bar \psi_-^T(z) & \sim & \left( \frac{z}{\alpha} \right)^{3/2} \left( \begin{array}{c} Q(\omega, {\bf k}) \\ 1 \end{array} \right) \eta^*, 
\label{aBC}
\eqa
where $\eta$ and $\eta^*$ are Grassmann numbers (not spinors) that we use to impose boundary data.
Once the second components of the spinors are chosen to be $\eta$ and $\eta^*$, 
the complex functions $P(\omega,k)$ and $Q(\omega,k)$ are uniquely determined from the boundary condition near the horizon.
It is noted that we could have chosen the boundary condition near $z=0$ in different ways. 
This is because the two components of the spinors decay in the same power\cite{MUCK}.
Different choices of boundary condition may correspond to different field theories on the boundary\cite{KLEBANOV}.
However, we note that Eq. (\ref{aBC}) is a natural choice for the following reasons.
First,  the vector $( 0, 1 )$ along 
which we impose boundary data is an eigenvector of $\sigma^z$, 
the generator of the rotation in the $x-y$ plane.
Therefore, this prescription is independent of momentum direction, which
 guarantees that the propagator that we will calculate below is invariant under the rotation. 
Second, once the second component of $\psi_-$ is chosen as boundary data, 
it is natural to choose the second component of $\bar \psi_-$ as boundary data.
This can be seen by turning on a small fermion mass which mixes $\psi_-$ and $\psi_+$,
 and identifying $\bar \psi_- \sim \psi_+^\dagger$.
Although not shown here, if we choose the $(1,0)$ component as our boundary data, 
which is another possible choice consistent with the above conditions, 
we obtain the same spectral function.

From the AdS/CFT dictionary, the Green's function of the fermion in the boundary theory is given by  
$G( \omega, {\bf k} ) = 
i \frac{ \partial^2 S[ \eta^*, \eta ] }{\partial \eta \partial \eta^* }$,  
where $S[ \eta^*, \eta ]$ is the gravity action evaluated for the saddle configuration of the spinor fields
which satisfy the boundary conditions, Eqs. (\ref{inBC}) and (\ref{aBC}).
The bulk spinor action in Eq. (\ref{action}) vanishes at saddle points.
Only the boundary term contribute to the action and we obtain the Green's function
\bqa
G( \omega, {\bf k} ) = 
i \left( P(\omega, {\bf k}) Q(\omega, {\bf k}) + 1 \right).
\eqa
The quantity of physical importance is the spectral function, $A(\omega,{\bf k}) = \lim_{\delta \rightarrow 0^{+}} Im G( \omega + i \delta, {\bf k})$ which 
measures how much spectral weight a fermion with momentum ${\bf k}$ has at energy $\omega$.

\begin{figure}[ht]
\centering
\includegraphics[width=0.5\textwidth]{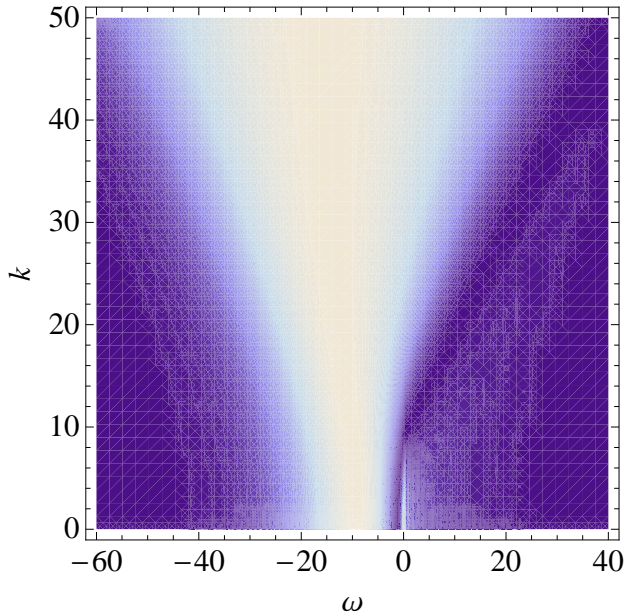} 
\caption{ 
The contour plot of the zero temperature spectral function as a function of energy and momentum for 
$q = -\sqrt{3}$ and $\alpha = 10$.
The darkest region represents the area with no spectral weight and
the brightest region, the highest value of the spectral function.
}
\label{fig:Aall}
\end{figure}

We numerically integrate the equation of motion (\ref{eom}) to obtain the spectral function as a function of $\omega$ and $k = \sqrt{k_x^2 + k_y^2}$.
Due to the rotational symmetry, $A(\omega,{\bf k})$ does not depend on  momentum direction.
In Fig. \ref{fig:Aall}, we show the spectral function at zero temperature.
For a large momentum, the spectral function as a function of energy shows a broad peak 
which is centered at a negative energy $\omega_0$.
The broad peak does not disperse significantly as momentum changes.
However, the width of the broad peak becomes larger as momentum increases
and the edges of the broad peak disperse as $\omega_{edge} \approx \pm k + \omega_0$.
There is no quasiparticle peak, which implies that 
the fermions are in a non-Fermi liquid state.

\bw
\begin{figure}[ht]
\centering
\subfigure[]{\label{fig:k3}\includegraphics[width=0.4\textwidth]{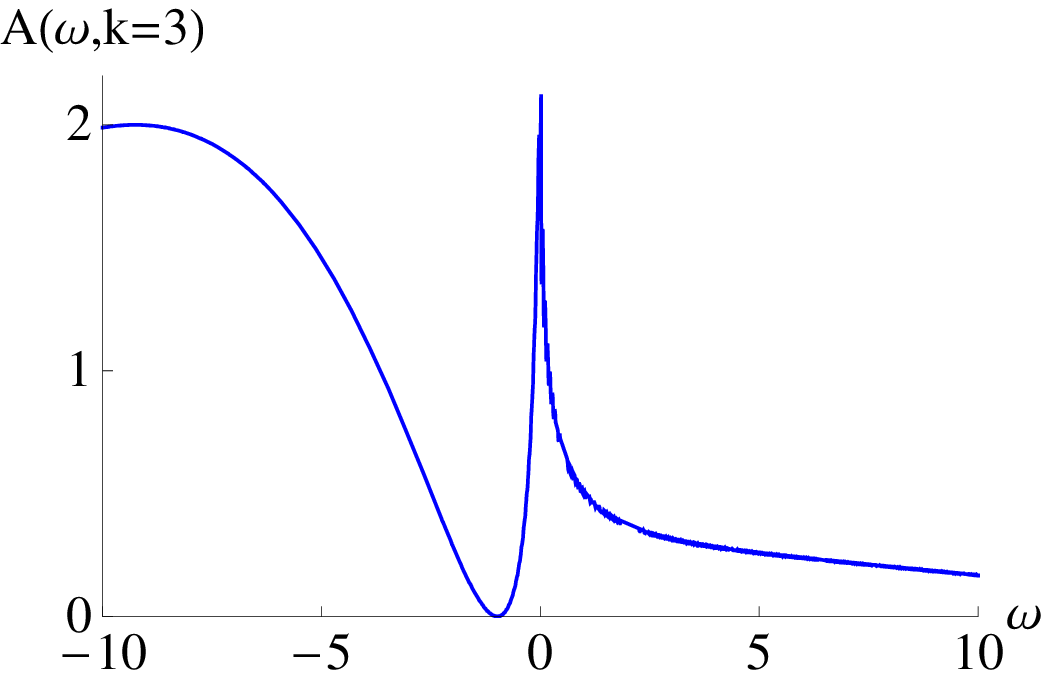}} 
\subfigure[]{\label{fig:k5}\includegraphics[width=0.4\textwidth]{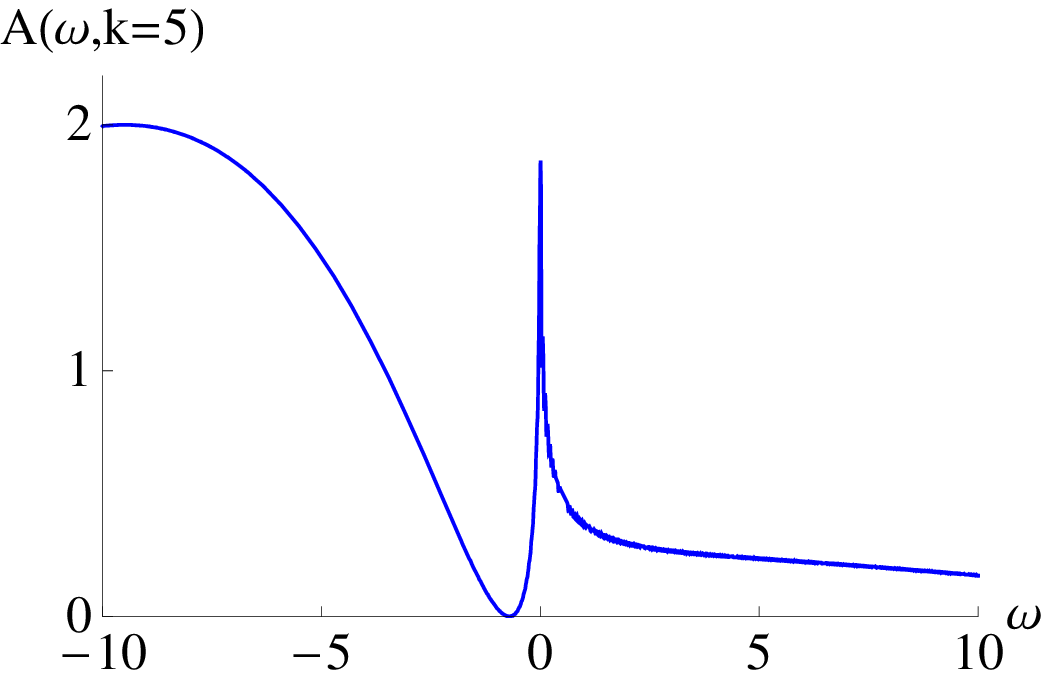}}
\subfigure[]{\label{fig:k7}\includegraphics[width=0.4\textwidth]{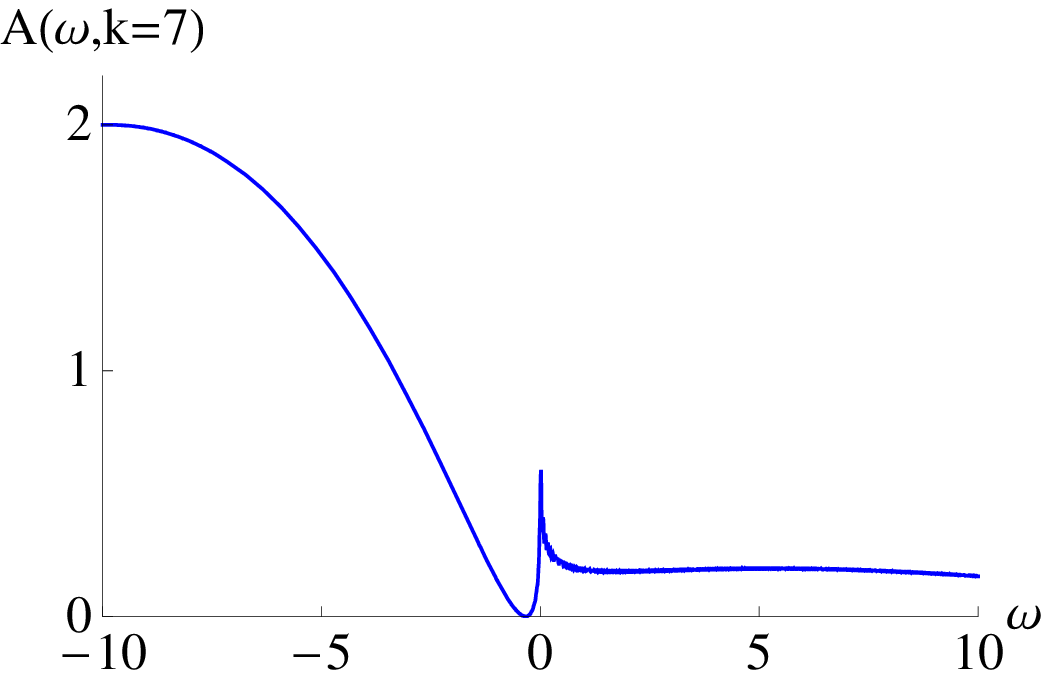}} 
\subfigure[]{\label{fig:k9}\includegraphics[width=0.4\textwidth]{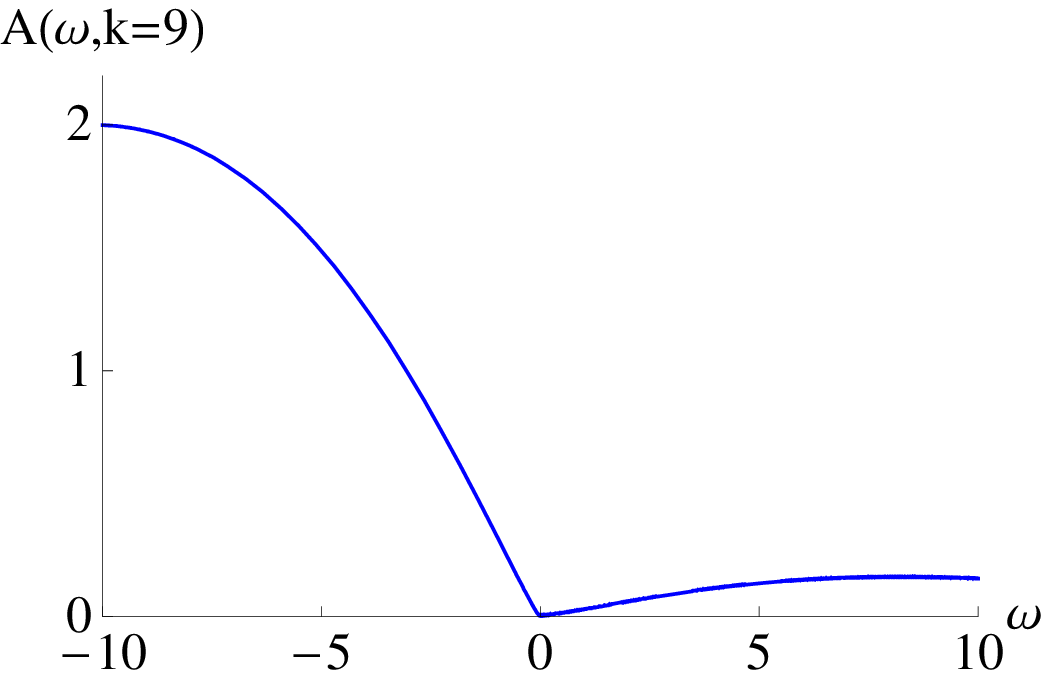}} 
\subfigure[]{\label{fig:k11}\includegraphics[width=0.4\textwidth]{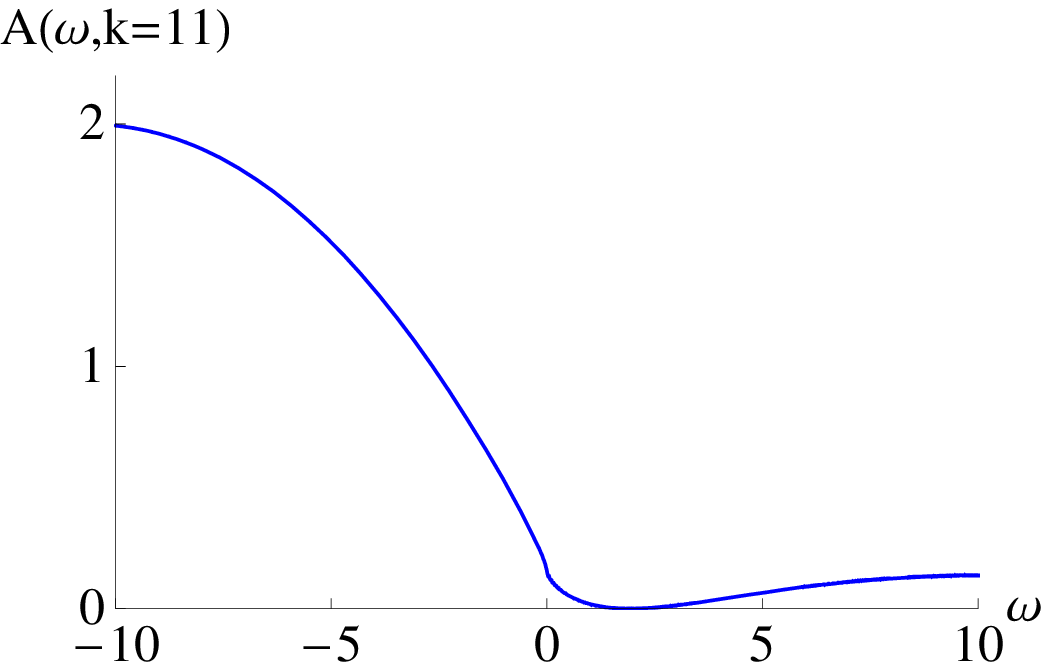}} 
\subfigure[]{\label{fig:k13}\includegraphics[width=0.4\textwidth]{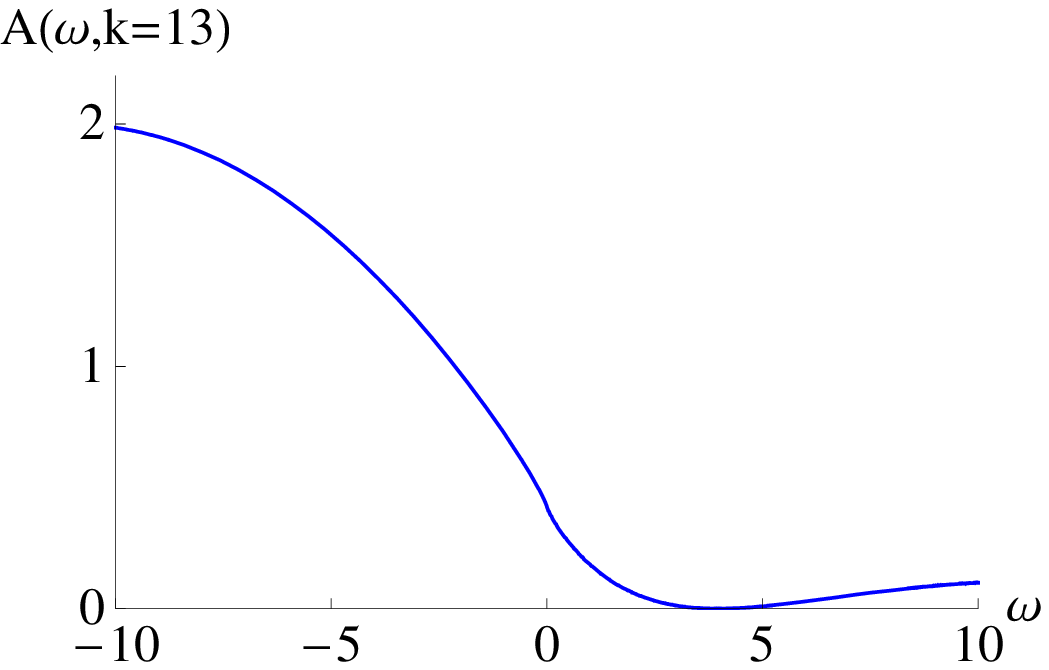}} 
\caption{ 
The energy distribution curves of the spectral function 
for $q = -\sqrt{3}$ and $\alpha = 10$ at momenta
(a) $k=3$, (b) $k=5$, (c) $k=7$, (d) $k=9$, (e) $k=11$ and (f) $k=13$.
}
\label{fig:Ak}
\end{figure}
\ew

Although there is no delta function peak, the spectral function shows sharp peaks at zero energy for momenta smaller than a critical momentum $k_c$.
To closely examine the low energy structure, we display the spectral function as a function of energy for fixed values of momentum in Fig. \ref{fig:Ak}.
Within the numerical accuracy, the sharp peak has an algebraic singularity at $\omega=0$.
We emphasize that this is not a quasiparticle peak.
As is shown in Fig. \ref{fig:Ak}, the zero energy peak is more pronounced at a smaller momentum
and the size of the peak decreases as momentum increases.
The critical momentum $k_c$ above which the zero energy peak disappears coincides with 
the momentum at which the edge of the broad peak cross the Fermi energy $\omega=0$.
Therefore, we interpret $k_c$ as Fermi momentum.
The most striking feature is that the algebraic singularities at zero energy exist 
for all momenta below the Fermi momentum.
Namely, all momentum points inside a two dimensional disk with 
$|{\bf k}| < k_c$ has the singular peak at zero energy.
We call the set of these momentum points a {\it critical Fermi ball}.
Although not shown here, the Fermi momentum and the absolute value of the energy of the broad peak increases as $\alpha$ increases.
If we switch the sign of $q$, the broad peak is centered at a positive energy.
However, the position of the critical Fermi ball does not change.

\begin{figure}[ht]
\centering
\includegraphics[width=0.5\textwidth]{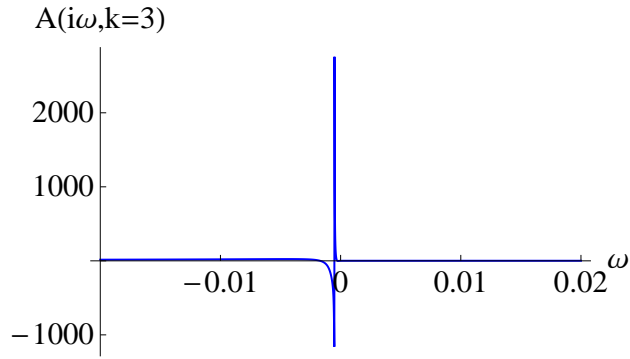} 
\caption{ 
The spectral function at $k=3$ as a function of imaginary frequency with the same parameters used in Fig. \ref{fig:Ak}.
}
\label{fig:Ak3imw}
\end{figure}

To show that there is true singularity at $\omega = 0$, the spectral function at a momentum below $k_c$ is shown as a function of imaginary frequency in Fig. \ref{fig:Ak3imw}.
Indeed, the spectral function has a strong singularity near $\omega=0$ along the imaginary axis.
A careful reader may note that the height of the peak is still finite and the position of the peak is slightly away from $\omega=0$.
These are artifacts originated from the fact that Eq. (\ref{eom}) has been numerically integrated over a range $[0, 1-\epsilon]$ with a small but nonzero $\epsilon$ to avoid the divergence in the equations at $z=1$.
As a smaller $\epsilon$ is used, the position of the peak moves to $\omega=0$ and the height of peak increases systematically.
This implies that the singularity at $\omega=0$ is genuine.

Unlike the Fermi liquid state or a non-Fermi liquid state with a critical Fermi surface\cite{CFS}
where low energy excitations exist only near a Fermi surface,
in the present non-Fermi liquid state all momentum points below the Fermi surface
are important at low energies.
Therefore we expect that the low energy properties of this state to be drastically different from   a Fermi liquid state or a non-Fermi liquid state with a critical Fermi surface.
For example, low temperature thermodynamic properties of this 2+1D non-Fermi liquid state with a critical Fermi ball will behave like a 3+1D critical Fermi surface.
In a sense, this `dimensional lift' is not surprising because the 2+1D non-Fermi liquid theory is described by the 3+1D gravity.

What would be the origin of the non-Fermi liquid behavior?
In the most trivial scenario, the non-Fermi liquid behavior can be caused by a composite nature of the fermion field.
If the fermion is a composite of weakly interacting fields, it will decay into multiple modes and the  spectral function will show a broad feature without a quasiparticle peak.
However, the sharp zero energy peak in the spectral function suggests that 
the fermion field is not a mere composite of weakly interacting fields.
The fermion is a rather well defined excitation at low energies 
irrespective of whether it is a fundamental or composite particle.
Then the non-Fermi liquid behavior can be due to strong interactions between the fermions.

Although the occurrence of the critical Fermi ball is somewhat counter-intuitive, one may understand it as a consequence of strong interactions.
Since the gravitational description is valid in the strong coupling limit, where the interaction energy scale is presumably larger than the Fermi energy, even those fermions which are deep inside the Fermi surface can participate in the low energy physics, overcoming the kinetic energy penalty.

\begin{figure}[ht]
\centering
\includegraphics[width=0.5\textwidth]{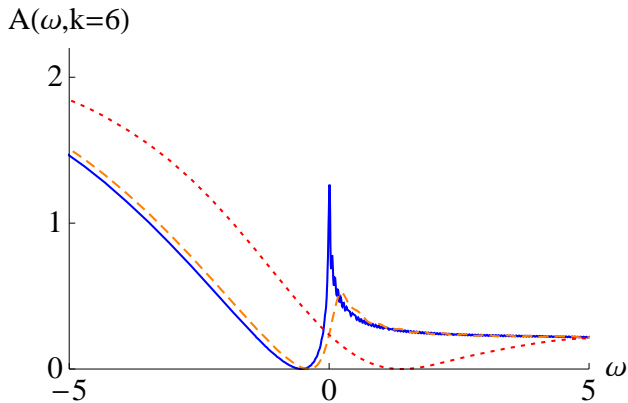} 
\caption{ 
Temperature dependence of the spectral function at $k=6$.
With a fixed $\alpha=10$, $q$ is changed to tune temperature to
$T=0$ (solid line), $T=\frac{1}{4 \pi}$ (dashed line) and $T=\frac{10}{4 \pi}$ (dotted line).
}
\label{fig:ATk}
\end{figure}

Until now, we have examined the zero temperature spectral function.
In Fig. \ref{fig:ATk}, we compare the spectral function at zero temperature and finite temperatures .
As expected, the singular zero energy peak is rounded at finite temperature due to thermal fluctuations.
If temperature is high enough, the sharp peak completely disappears.
At finite temperatures, the Fermi momentum is not sharply defined, but the position of the broad peak is not sensitive to temperature.

In summary, we solved the Dirac equation in a charged black hole background to extract a fermionic spectral function of a  2+1 dimensional strongly coupled field theory at finite chemical potentials.
The spectral function revealed a critical Fermi ball in the momentum space where all momentum points inside the Fermi ball are critical.

In the future, it would be interesting to study physical properties of the critical Fermi ball in more details, such as possible instabilities and thermodynamic/transport properties. 
Due to the presence of extensive gapless modes, it is expected that there are infinitely many singular channels of particle-hole and particle-particle excitations with different momenta which compete with each other.

I thank Sean Hartnoll, Pavel Kovtun, Subir Sachdev and Arkady Tseytlin for helpful comments.
This work has been supported by NSERC.

\end{document}